\documentclass[a4paper,aps,prd,twocolumn,noshowpacs,preprintnumbers,nofootinbib]{revtex4-1}

\usepackage[utf8x]{inputenc}
\usepackage{epsfig,amssymb,amsmath,psfrag}

\usepackage{etex}
\usepackage{pstricks-add}

\usepackage{bm}
\usepackage{epic}
\usepackage{graphicx}
\usepackage{color}
\usepackage{hyperref}
\usepackage{young}
\usepackage{multirow}
\usepackage{fancybox}

\usepackage[paperwidth=210mm,paperheight=297mm,centering,hmargin=1.75cm,vmargin=2.5cm]{geometry}

\def \be  {\begin{equation}}
\def \ee  {\end{equation}}
\def \ba  {\begin{eqnarray}}
\def \ea  {\end{eqnarray}}
\def \baa {\begin{eqnarray*}}
\def \eaa {\end{eqnarray*}}
\def \bb  {\begin {thebibliography} }
\def \eb  {\end{thebibliography}}
\def \lab #1 {\label{#1}}

\newcommand{\beq}{\begin{equation}}
\newcommand{\eeq}{\end{equation}}
\newcommand{\beqa}{\begin{eqnarray}}
\newcommand{\eeqa}{\end{eqnarray}}

\begin{document}

\preprint{HU-EP-13/16}

\title{The $1/2$ BPS Wilson loop in ABJM theory at two loops}

\author{Marco S. Bianchi${}^{1}$}
\author{Gaston Giribet${}^{2}$}
\author{Matias Leoni${}^{2}$}
\author{Silvia Penati${}^{3}$}

\affiliation{
${}^1${
Institut f\"ur Physik,
Humboldt-Universit\"at zu Berlin,
Newtonstra{\ss}e 15, 12489 Berlin, Germany}\\
${}^2${Physics Department, FCEyN-UBA \& IFIBA-CONICET\
Ciudad Universitaria, Pabell\'on I, 1428, Buenos Aires, Argentina}\\
${}^3${Dipartimento di Fisica, Universit\`a degli studi di Milano--Bicocca and
  INFN, Sezione di Milano--Bicocca, Piazza della Scienza 3, I-20126 Milano, Italy}\\
{\tt marco.bianchi@physik.hu-berlin.de, gaston@df.uba.ar, leoni@df.uba.ar,
    silvia.penati@mib.infn.it}}
\begin{abstract}
We compute the expectation value of the $1/2$ BPS circular Wilson loop in ABJM theory at two loops in perturbation theory.
The result shows perfect agreement with the prediction from localization and the proposed framing factor.
\end{abstract}

\maketitle

\section{Introduction}
\label{sec:introduction}

One of the most interesting instances of AdS/CFT\ correspondence is the
duality between string theory on AdS$_{4}\times \mathbb{CP}^{3}$ background and
the $\mathcal{N}=6$ superconformal extension of Chern-Simons theory (CS)
constructed in \cite{ABJM, ABJ}, hereafter referred to as ABJM theory. The
precise statement of the duality is that type IIA string theory on AdS$%
_{4}\times \mathbb{CP}^{3}$ is dual to the $\mathcal{N}=6$ CS theory coupled
to bifundamental matter, with gauge group $U(N)\times U(M)$, in the limit of large $N,$ $M$
and large CS\ level $k$, with $\lambda =N/k$ and $\widehat{\lambda }=M/k$
fixed\footnote{Precisely, the theory is referred to as ABJM when 
$M=N$ and ABJ when $M \neq N$, but we will not make distinction
here.}. 
In the last four years, this particular realization of the AdS/CFT
correspondence has been extensively investigated, and different observables
of the theory, such as scattering amplitudes \cite{ABL}--\cite{yutin} 
and Wilson loops (WL) \cite{Rey}--\cite{Bianchi:2011rn} were
studied. In this paper, we will be concerned with the computation of the
expectation value of WL operators in ABJM theory.

Among the gauge theory observables WL are of particular importance. This is
due to the fact that, on the one hand, WL encode important information
about gauge theory, such as the interaction potential
of colored particles and Schwinger pair production probability. On the other
hand, in the context of holography, WL happen to be the gauge theory duals to
fundamental string states in AdS spaces. In the special case of ABJM
theory, it containing a CS contribution, WL are natural observables to look
at. Moreover, the evaluation of WL allows to investigate the
intriguing relation that seems to exist between perturbative results of the
ABJM theory and the $\mathcal{N}=4$ four-dimensional super Yang-Mills (SYM) (see for instance \cite{Rey}).

Supersymmetric circular WL\ in ABJM\ theory were first discussed in Refs$.$ \cite{Drukker, Chen, Rey}, where operators preserving $1/6$ of the supersymmetry were constructed as the holonomy of the connections
${\cal A} = A_\mu \dot{x}^\mu - \frac{2\pi i}{k}  |\dot{x}| {\mathcal M}_J^{\; I} C_I \bar{C}^J$ and $\hat{\cal A} = \hat{A}_\mu 
\dot{x}^\mu - \frac{2\pi i}{k}  |\dot{x}| \hat{{\mathcal M}}_J^{\; I}   \bar{C}^J C_I$, combining the gauge fields $A$ and $\hat A$ with the 
bifundamental scalars $C$ and $\bar C$ through proper matrices ${\mathcal M}=\hat {\mathcal M}={\rm diag}(-1,-1,1,1)$.

Such operators are direct generalizations of the
configurations previously studied in $\mathcal{N}=4$ SYM \cite{Maldacena,Rey:1998ik,Zarembo:2002an}, and they also
exist in CS theories with less supersymmetry. Of particular importance to
our discussion are the $1/6$ BPS\ operators studied in Ref$.$ \cite{Rey},
where linear combinations of WL transforming oppositely under time-reversal
were considered and perturbative computations for their expectation values
were performed.

A year after $1/6$ BPS\ WL\ operators were found, Kapustin et al. discussed
in \cite{Kapustin} the computation of WL expectation values in CS theories with matter
by resorting to localization techniques \cite{Pestun:2007rz}. In Ref$.$ \cite{MarinoPutrov}, the
matrix model constructed in \cite{Kapustin} was used to compute exact expressions for the expectation value
of the $1/6$ BPS\ WL operators in ABJM in the large $N$ limit, providing in this way
an interpolating function between the weak and the strong coupling regimes
of the theory. 

Simultaneously, in Ref$.$ \cite{DrukkerTrancanelli}, a WL
operator that preserves $1/2$ of the supersymmetry in ABJM was found. As
shown in \cite{DrukkerTrancanelli}, the different couplings of this $1/2$
BPS\ operator to the matter fields of the theory can be accommodated in a
single superconnection ${\cal L}$ for the supergroup $U(N|M)$
\begin{equation}
\label{eq:supermatrix}
{\cal L}(\tau) \, = \, \left( \begin{array}{ccc}
{\cal A} & -i\, \sqrt{\frac{2\pi}{k}}\, |\dot{x}|\, \eta_I\, \bar{\psi}^I \\
 -i\, \sqrt{\frac{2\pi}{k}}\,  |\dot{x}|\, \psi_I\, \bar{\eta}^I & \hat{\cal A}  \end{array}\right)   ,
\end{equation}
whose holonomy gives the WL
\begin{equation}
\label{eq:WL}
W_{1/2}[C]= \frac{1}{N+M}\, {\rm Tr}\, P\, \exp{
\left( - i \int_C d\tau\, {\cal L}(\tau) \right) } 
\end{equation}
${\cal A}$ and $\hat{\cal A}$ are the same combinations of gauge and scalar fields as for the 1/6 BPS case, although with a different matrix 
${\mathcal M}={\rm diag}(-1,1,1,1)$; $\eta_I$ and $\bar \eta^I$ are commuting spinors controlling the coupling to the bifundamental fermions $\psi$ 
and $\bar \psi$, whose 
supersymmetry preserving form for a circular WL is spelled out in \cite{DrukkerTrancanelli}.

It was observed in \cite{DrukkerTrancanelli} that the $1/6$ BPS\ and the $1/2$ BPS\ circular WL operators
happen to belong to the same cohomology class under the supercharge used for localization. This results in an equivalence
between their corresponding expectation values. In particular, adapting the matrix model of \cite{Kapustin} to the superconnection
representation, this enables to
derive a prediction for the expectation value of the $1/2$ BPS operator. Here, we present a perturbative test of this prediction. 

In Section \ref{sec:1} we review the matrix model calculation and the prediction for the weak coupling expectation value, particularly in the case of the $1/2$ BPS WL operator. In Section \ref{sec:2} 
we present the explicit computation of the $1/2$ BPS operator in planar perturbation theory, up to two loops. This includes the novel fermionic 
diagrams, which turn out to be the most technically involved contributions. Finally, in Section \ref{sec:conclusions} we compare the results we obtain in perturbation theory with those coming from localization. We find perfect matching, once the result from localization gets mapped to framing zero by suitably removing the framing phase.

\section{Localization}
\label{sec:1}

In Ref$.$ \cite{Kapustin}, by means of localization techniques, it was shown that the path integral of the supersymmetric CS theory reduces to a non-Gaussian matrix model. This matrix model yields the following expression for the
partition function 
\begin{eqnarray}
\mathcal{Z}&=&\int \prod_{a=1}^{N}d\lambda _{a} \ e^{i\pi k\lambda
_{a}^{2}}\prod_{b=1}^{M}d\hat{\lambda }_{b} \ e^{-i\pi k\widehat{%
\lambda }_{b}^{2}} \times \label{Z}\\
&& \frac{\prod_{a<b}^{N}\sinh ^{2}(\pi (\lambda
_{a}-\lambda _{b}))\prod_{a<b}^{M}\sinh ^{2}(\pi (\hat{\lambda 
}_{a}-\hat{\lambda }_{b}))}{\prod_{a=1}^{N}\prod_{b=1}^{M}\cosh ^{2}(\pi (\lambda _{a}-\hat{\lambda }_{b}))}\nonumber 
\end{eqnarray}

The evaluation of expectation values of $W_{1/6}[C]$ and $\hat{W}%
_{1/6}[C]$ amounts to inserting in (\ref{Z}) the following contributions%
\begin{equation}
w_{1/6}=\frac{1}{N}\sum_{a=1}^{N}e^{2\pi \lambda _{a}}\quad \text{%
and}\quad \hat{w}_{1/6}=\frac{1}{M}\sum_{a=1}^{M}e^{2\pi 
\hat{\lambda }_{a}}  \label{Wsum}
\end{equation}
where $w_{1/6}$ and $\hat{w}_{1/6}$ correspond to the $U(N)$ and
$U(M)$ pieces, respectively.

As said above, it was observed in \cite{DrukkerTrancanelli} that the $1/6$ BPS and the $1/2$ BPS circular WL operators
belong to the same cohomology class, and this yields an equivalence
between the corresponding expectation values. In fact, computing the $1/2$ BPS operator in terms of the matrix model
amounts to plugging in (\ref{Z}) the operator
\begin{equation}
w_{1/2}=\frac{1}{N+M}\left( \sum_{a=1}^{N}e^{2\pi \lambda _{a}}+\sum_{a=1}^{M}e^{2\pi \hat{\lambda }_{a}} \right) \label{sumW}
\end{equation}
From equations (\ref{Wsum}) and (\ref{sumW}), we observe the relation $(N+M)\,w_{1/2}=N\,w_{1/6}+M\,\widehat{w}_{1/6}$. Therefore, 
at the level of the expectation 
values, the relation between the $1/6$ BPS\ and the $1/2$ BPS operators can be
expressed by \cite{DrukkerTrancanelli}
\begin{equation}
\left\langle W_{1/2}[C]\right\rangle _{\text{f}=1} =\frac{N\big\langle
W_{1/6}[C]\big\rangle _{\text{f}=1} +M\big\langle \hat{W}_{1/6}[C]\big\rangle _{\text{f}=1} }{N+M}
\label{eq:W}
\end{equation}
where $\left\langle W_{1/6}[C]\right\rangle _{\text{f}=1} $ and $\langle \hat{W}%
_{1/6}[C]\rangle _{\text{f}=1} $ represent the expectation values of the $1/6$ BPS\ operators
corresponding to the $U(N)$ and $U(M)$ pieces, respectively. 
Since coming from localization, identity (\ref{eq:W}) has to be understood at the nontrivial framing one, and this is precisely 
what the subindices 
${\text{f}=1}$ in (\ref{eq:W}) refer to. Framing \cite{Witten, GMM, Labastida} is the choice 
of a normal vector field along the contour $C$, defining a nearby path which allows for a point splitting regularization of the self--linking number of a 
knot in a topologically invariant manner.
For pure Chern--Simons this procedure has been shown to provide sensible results \cite{GMM}, since the WL expectation values do not depend on the particular 
choice of the framing contour, but just on its linking number with the original path. In order for framing to be compatible with localization, it has 
to respect supersymmetry, and this in turn imposes that the computation be performed at framing one \cite{Kapustin}. 
Then, using the matrix model one can compute the framing one expectation value of these
operators at weak coupling in the planar limit and obtain \cite{Drukker:2010nc}
\begin{align}
\label{onehalfWL}
& \left\langle W_{1/2}[C]\right\rangle _{\text{f}=1} = 1+\frac{i\pi }{k}(N-M) \notag \\
&~~~~~~~~~~ -\frac{\pi ^{2}}{3k^{2}}(2N^{2}+2M^{2}-5NM)+  
\mathcal{O}(1/k^{3})
\end{align}
In order to compare with a perturbative computation performed at framing zero, one has to identify and remove the framing phase. This was carried out successfully for the $1/6$ BPS WL's for which the prescriptions \cite{Kapustin}
\begin{align}
\label{framing0}
& \big\langle W_{1/6}[C]\big\rangle  _{\text{f}=0} = e^{-\frac{i\pi}{k}N} \, \big\langle W_{1/6}[C]\big\rangle  _{\text{f}=1} \notag \\
& \big\langle \hat{W}_{1/6}[C]\big\rangle  _{\text{f}=0} = e^{\frac{i\pi}{k}M} \, \big\langle \hat{W}_{1/6}[C]\big\rangle  _{\text{f}=1} 
\end{align}
provide results which match the perturbative evaluation \cite{Rey, Drukker, Chen}.  

For the $1/2$ BPS WL, the framing factor has been identified in (\ref{onehalfWL}) as \cite{Drukker:2010nc} 
\begin{align}
\label{onehalfWL2}
& \left\langle W_{1/2}[C]\right\rangle _{\text{f}=1} = e^{\frac{i\pi }{k}(N-M)} \left\langle W_{1/2}[C]\right\rangle _{\text{f}=0} =   e^{\frac{i\pi }{k}(N-M)}  \notag \\
& \times \left[ 1 -\frac{\pi^2}{6\, k^2} \left( N^2+M^2 - 4 NM \right)+ \mathcal{O}(1/k^3) \right]
\end{align}
We will prove that the square bracket is indeed the perturbative result at framing zero.

\section{Perturbative computation}
\label{sec:2}

In this Section, we present the perturbative evaluation of the circular $1/2$ BPS WL in ABJM, up to two loops in the planar limit. Since such a computation is quite involved, requiring the regularization and the calculation of intricate trigonometric integrals, here we simply quote the main results referring to \cite{BGLP} for details on the calculation and the techniques used.
 
We start from the expression (\ref{eq:WL}) for the WL where $C$ is a circle parametrized as $x^{\mu}_i = (0, \cos{\tau_i}, \sin{\tau_i})$. 
In ordinary perturbation theory the evaluation is performed as usual by Taylor expanding the exponential of the superconnection (\ref{eq:supermatrix}) and taking the expectation value by Wick contracting the fields. Since we are interested in the two--loop quantum corrections, it suffices to expand it up to the fourth order. In this process we get purely bosonic contributions from the diagonal part of the $U(N|M)$ super-matrix (\ref{eq:supermatrix}), purely fermionic contributions from the off--diagonal blocks and mixed contributions from the mixing of the two.

In order to deal with potentially divergent diagrams, we use dimensional regularization with dimensional reduction (DRED), which has been proven to preserve gauge invariance and supersymmetry of Chern--Simons theories up to two loops \cite{Chen:1992ee}. This amounts to perform all tensor manipulations in three dimensions before promoting loop integrals to $D= 3-2\epsilon$. In particular,  care is required when contracting 3d metric tensors coming from Feynman rules with D-dimensional metric tensors coming from tensor integrals. Moreover, in order to avoid potential problems arising from contracting 3d epsilon tensors (coming from gauge propagators and cubic vertices) with tensorial loop integrals we use helpful identities for eliminating $\varepsilon_{\mu\nu\rho}$ tensors before analytically extending the integrals to D-dimensions. 

The integrals generally converge in the complex half--plane defined by some critical value of the real part of the regularization parameter $\epsilon$. Using techniques that will be presented in \cite{BGLP}, we have been able to compute them analytically for any complex value of $\epsilon$. They turn out to be expressible in terms of hypergeometric functions. In the regime where they converge we have successfully tested our results numerically. In a neighborhood of $\epsilon=0$, we have evaluated them by analytically continuing the hypergeometric functions and expanding the result up to finite terms.

At one--loop, neglecting tadpoles and diagrams vanishing due to the antisymmetry of the $\varepsilon$ tensor,  the only non--trivial Feynman diagram comes from a fermion exchange. The corresponding integral can be easily evaluated and turns out to be subleading in $\epsilon$. Thus, $\left\langle W_{1/2}[C]\right\rangle^{(1)} _{\text{f}=0} =0$, in line with the prediction from localization when the framing factor (\ref{onehalfWL2}) is removed. 

At two loops, neglecting diagrams which vanish identically because of the antisymmetry of the $\varepsilon$ tensor, we are left with purely bosonic diagrams coming from contracting the diagonal terms in (\ref{eq:supermatrix}) plus two extra diagrams where exchanges of fermions appear.  

The diagrams arising from the contraction of the diagonal blocks are already present in the evaluation of $1/6$ BPS WL. Although the definition of the 
${\mathcal M}$ matrix governing the coupling to the scalars is different in the two cases, only 
the trace of ${\mathcal M}^2$ enters at this order, which is exactly the same. Therefore, borrowing the results for the $1/6$ BPS case \cite{Rey, Drukker, Chen}, the sum of the relevant contributions from the upper diagonal block reads
\begin{equation}
\label{bosons}
\raisebox{-5.5 mm}{\includegraphics[scale=0.25]{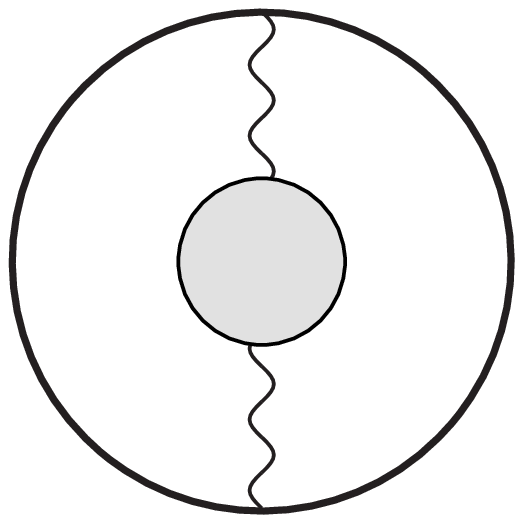}}
+
\raisebox{-5.5 mm}{\includegraphics[scale=0.25]{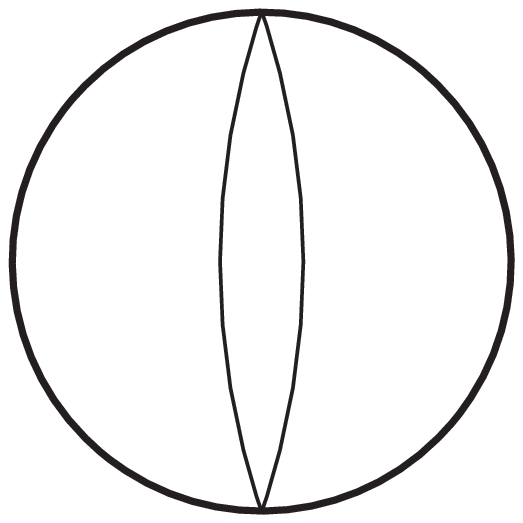}}
+
\raisebox{-5.5 mm}{\includegraphics[scale=0.25]{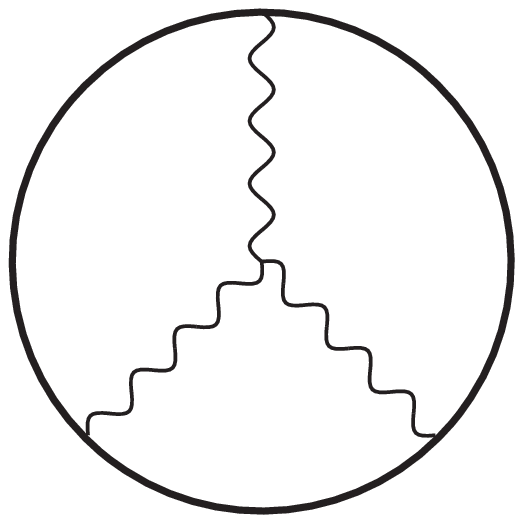}}
\, =\, \frac{N^2 M - \tfrac16 N^3}{M+N}\, \frac{\pi^2}{k^2}
\end{equation}
Similarly, the contribution from the lower diagonal block is obtained upon exchanging $M \leftrightarrow N$.
 
Two additional contributions involving exchange of fermions need be considered, one with a fermionic ladder and one with a gauge/fermion cubic interaction. Potentially, also a diagram with the exchange of a one--loop corrected fermion propagator should be considered, but it vanishes  when taking the trace of the superconnection exponential.  

The fermionic ladder is straightforwardly computed by calculating the relevant integrals over the WL parameters. There are two contributions depending 
on which pairs of fermions are contracted. Even though they are individually divergent, their sum is finite and up to order ${\cal O}(\epsilon)$
it is given by
\begin{equation}
\label{ladder}
\raisebox{-5.5 mm}{\includegraphics[scale=0.25]{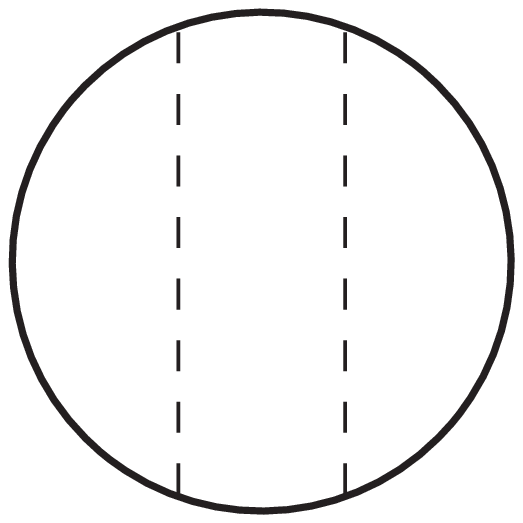}}
+
\raisebox{-5.5 mm}{\includegraphics[scale=0.25]{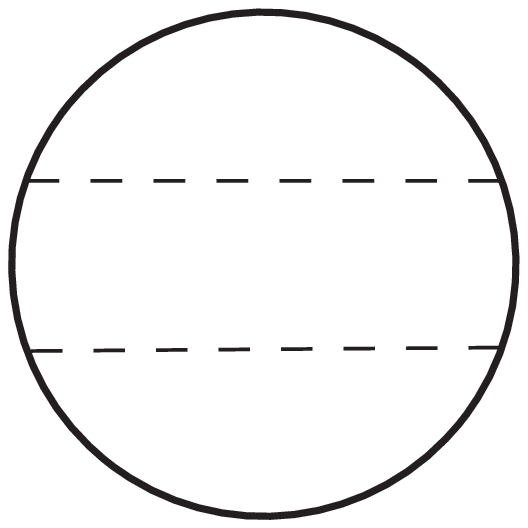}}
~ = ~ \frac32 \pi^2\, \frac{M N}{k^2}
\end{equation}
The vertex graph is troublesome, since it involves both an integration over the WL parameters and a space--time integral from the internal interaction vertex. By carrying out the algebra of the diagram we end up with a linear combination of integrals of the form
\begin{equation}
\partial_{1}^{\mu} \,  \partial_{2}^{\nu} \,  \partial_{3}^{\rho} \, \int \frac{d^{3-2\epsilon}x}{[(x-x_1)^2]^{\frac12-\epsilon} [(x-x_2)^2]^{\frac12-\epsilon} [(x-x_3)^2]^{\frac12-\epsilon}} \label{derivadas}
\end{equation}
where $\partial_i^\mu \equiv \frac{\partial}{\partial x_{i \mu}}$. In some terms the indices carried by the derivatives are (partially) contracted among themselves, whereas in other terms they are contracted with external tensorial structures. 

Whenever two of the indices are contracted the computation simplifies, since we can exploit the D-dimensional Green equation for the propagators and get rid of the space--time integral.
After collecting all such pieces and performing the parametric integrals we find a divergent contribution 
\begin{equation}
\raisebox{-5.5 mm}{\includegraphics[scale=0.25]{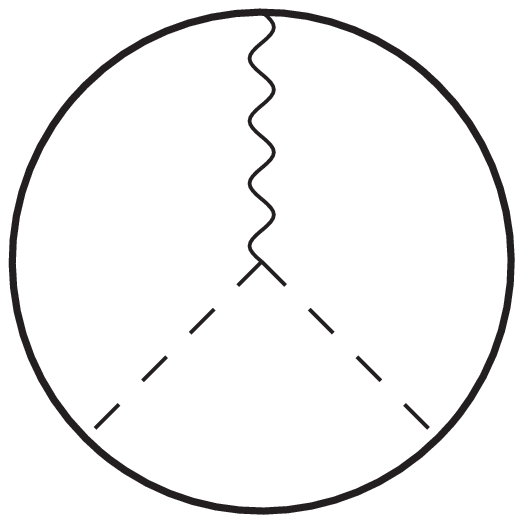}}
~~ \bigg|_{c} ~ = ~ \pi^{2+2\epsilon}\, \frac{MN}{2 k^2} \left( -\frac{1}{2 \epsilon }- 1 - \gamma_E - 4 \log 2 \right)
\end{equation}
where $|_{c}$ refers to contributions that involve integral (\ref{derivadas}) with the indices contracted. 

In the remaining terms the space--time integration is unavoidable. Still, we were able to carry it out and the final contribution reads
\begin{equation}
\raisebox{-5.5 mm}{\includegraphics[scale=0.25]{2LfermionMB.eps}}
~ \bigg|_{u} ~ = ~ \pi^{2+2\epsilon}\, \frac{M N}{2 k^2}
\left(\frac{1}{2\epsilon}-3+\gamma_E+4\log 2 \right)
\end{equation}
where in this case $|_{u}$ stands for contributions that involve integral (\ref{derivadas}) with the indices uncontracted. Summing the contracted and uncontracted parts of the vertex diagram we can ascertain that the divergence cancels out, as well as all unpleasant constants. The final result from this diagram, up to  ${\cal O}(\epsilon)$ terms, reads
\begin{equation}
\label{vertex}
\raisebox{-5.5 mm}{\includegraphics[scale=0.25]{2LfermionMB.eps}} ~ = ~ - 2 \pi^2 \, \frac{MN}{k^2}  
\end{equation}
  
Therefore, summing the contribution (\ref{bosons}) plus its analogous from $M \leftrightarrow N$ with the contributions (\ref{ladder}) and (\ref{vertex}), the expectation value of the $1/2$ BPS WL at two loops in the planar limit is
\begin{equation}
\left\langle W_{1/2} [C] \right\rangle^{(2)} _{\text{f}=0} = - \frac{\pi^2}{6\, k^2} \left[ N^2+M^2 - 4\, NM \right] 
\end{equation}
Notably, this equals the prediction (\ref{onehalfWL2}) from localization once the phase factor $e^{\frac{i\pi }{k}(N-M)} $ has been removed.

\section{Conclusions}
\label{sec:conclusions}

Up to two loops, we have determined the $1/2$ BPS circular Wilson Loop in the ABJM theory.
We have proved that the perturbative result perfectly matches the weak coupling prediction from localization, once the framing--one factor has been removed. 

As already mentioned,  in the $W_{1/6}[C]$ case the prescription for relating framing--one and framing--zero results by rescaling with a $e^{i\pi N/k}$ factor was supported by  matching the localization result with known results from perturbation theory. In the 
$\left\langle W_{1/2}[C]\right\rangle$ case, while the localization result at framing one has been easily inferred using identity (\ref{eq:W}), the prescription (\ref{onehalfWL}) for removing the framing factor, even if natural, was lacking a direct confirmation from a perturbative result.    
Here, we have provided the perturbative confirmation. 

This calculation is technically involved, and this is due to the fact that in contrast
to the $1/6$ \ BPS\ operators, which couple only to the gauge
fields and the scalars of the theory, the $1/2$  BPS operator
also couples to the fermions, which are in the bifundamental representation of $U(N)\times U(M)$.  
This introduces notable technical difficulties. 
In particular, individual contributions from fermionic diagrams are divergent at two loops and only in the end all singularities cancel out. Therefore one has to introduce a regularization for the integrals. This is a delicate point because finite terms are affected by the choice of regularization scheme.
It turns out that a careful application of DRED yields the correct result.
Dimensionally regulating integrals also sources technical issues, since it requires solving them analytically for any value of the parameter $\epsilon$, which is a hard task.
The details of such computations will be reported in a future publication \cite{BGLP}.

The techniques we used may be applied to the perturbative analysis of other Wilson loop operators.
In particular,  it would be interesting to extend the present computation to other supersymmetric Wilson loops in ABJM, such as the generalizations proposed in \cite{Griguolo,Cardinali}, for which an exact result from localization has not been derived yet.

\section{Acknowledgements}

We thank L. Griguolo, J. Maldacena and D. Seminara for very useful discussions.  
The work of MB has been supported by the Volkswagen-Foundation.
The work of GG and ML has been supported by the research project CONICET PIP0396.
The work of SP has been supported in part by INFN, MIUR--PRIN contract 2009--KHZKRX and MPNS--COST Action MP1210 "The String Theory Universe".

\end{document}